\documentclass[]{elsarticle}

\def\lsim{\mathrel{\rlap{\lower4pt\hbox{\hskip1pt$\sim$}}
    \raise1pt\hbox{$<$}}}                
\def\gsim{\mathrel{\rlap{\lower4pt\hbox{\hskip1pt$\sim$}}
    \raise1pt\hbox{$>$}}}                
    
\usepackage{amsmath}
\usepackage{amssymb}
\usepackage{amsfonts}
\usepackage{subfigure}
\usepackage{appendix}
\usepackage{listings}
\usepackage{fancyhdr}
\usepackage{a4wide}

\usepackage{graphicx}
\usepackage{color}
\usepackage{geometry}
\def\square{\vcenter{\vbox{\hrule height.4pt
          \hbox{\vrule width.4pt height8pt
          \kern8pt\vrule width.4pt}\hrule height.4pt}}}

\newcommand{\beq}{\begin{equation}}
\newcommand{\eeq}{\end{equation}}
\newcommand{\bqa}{\begin{eqnarray}}
\newcommand{\eqa}{\end{eqnarray}}

\begin{document}

\title{Functional renormalization group at finite density and Bose
condensation}
\date{\today}

\author{Eirik E. Svanes}

\ead{e.svanes1@physics.ox.ac.uk}
\author{Jens O. Andersen}
\ead{andersen@tf.phys.ntnu.no}
\address{Department of Physics, Norwegian University of Science
and Technology, H{\o}gskoleringen 5, N-7491 Trondheim, Norway}

\begin{abstract}
We discuss the functional renormalization group 
and pion condensation in the presence of a
finite isospin chemical potential $\mu_I$. 
We calculate the phase diagram as 
function of temperature $T$ and $\mu_I$.
While the exact effective average action is invariant under
certain gauge transformations, the effective action in the 
local-potential approximation is not.
As a consequence, the critical chemical potential $\mu_I^c$ for Bose-Einstein
condensation at $T=0$
is no longer equal to the mass of the condensing mode.
We discuss possible solutions to this problem.

\end{abstract}
\begin{keyword}
Pion condensation, functional renormalization group, finite-temperature
field theory
\end{keyword}
\maketitle

\section{Introduction}
One of the goals of heavy-ion collisions is to create
energy densities and temperatures 
high enough to create a quark-gluon plasma and 
to study part of the phase diagram of 
quantum chromodynamics (QCD).
The attempts to understand the phase structure of
QCD, together with the 
sign problem at nonzero baryon chemical potential $\mu_B$,
have triggered interest in several QCD-like theories.
These include QCD at nonzero isospin density~\cite{iso}, QCD
with adjoint quarks, and two-color QCD~\cite{twocolor}. A common
feature of these theories is that they are free of the sign
problem. On one hand this admits their straightforward
simulation using lattice Monte-Carlo techniques
and serves as a check of these methods against model-independent
predictions of chiral perturbation theory. On the other
hand, it allows a direct test of various model calculations 
at high temperature and/or density where chiral perturbation
theory is not applicable. Hence the study of
QCD-like theories contributes to our understanding of
the physics of strongly-coupled gauge theories at nonzero
temperature and density.

In the present paper, we consider the simpler problem
of pion condensation in the presence of
a finite chemical potential $\mu_I$ for isospin 
using the linear sigma model. 
Lattice simulations~\cite{lat1,lat2,lat3} 
suggest that there is a deconfinement transition
of pions at high temperature and low density, and Bose-Einstein condensation of
charged pions at high isospin density and low temperature. In fact, 
the deconfinement
transition and the transition to a charged pion condensate seem to coincide.
The deconfinement transition is found by measuring the Polyakov loop and the
measurements show a sharp increase (indicating deconfinement) at approximately 
the same temperature as the onset of pion condensation.
The ground state energy and equation of state were studied on the lattice
in~\cite{savage1,savage2}.
Pion condensation at finite $\mu_I$ has also been studied 
using chiral perturbation theory~\cite{iso,split,loewe}, 
ladder QCD~\cite{barducci}, the chiral quark model~\cite{jako,herpay}, 
the linear sigma model~\cite{mat,kapusta,weldon,heman2,jens1,jenstomas,shu},
Dyson-Schwinger equations~\cite{Dyson},
the Nambu Jona-Lasinio (NJL) 
model~\cite{heman2,ravagli,ebert1,ebert2,lawley,zhuang,jenslars1,abuki}, 
and Polyakov-loop NJL (PNJL) models~\cite{zhangliu,abuki2}. 
In Refs.~\cite{ebert2,jenslars1,abuki,abuki2}, the effects of charge neutrality
were also investigated.                
In contrast to the lattice simulations, 
the PNJL model suggests that deconfinement and the onset of Bose-Einstein 
condensation are two different transitions. Finally, the possibility
of a BEC-BCS crossover from a Bose condensate to superconducting state has 
been investigated~\cite{heman3}. Note that this
is not a phase transition since there is no global
order parameter that distinguishes
the crossover. Rather, it is a qualitative change from 
a pion condensate of
tightly bound quarks
to weakly bound Cooper pairs as the isospin density increases. 

The model calculations are typically of mean-field
type and going beyond mean field is the next step in the study of these models.
In order to do so we apply the functional renormalization group 
(FRG)~\cite{wetterich,review1,review2, review3,review4,review5,review6,review6p,review7}.
The paper is based on Ref.~\cite{eirik}.
The functional renormalization group
is a nonperturbative method, with a wide
range of applicability. It is an alternative, but equivalent formulations
to Wilson's ideas from the 1970s.
It has been used successfully to
calculate critical exponents for phase transitions~\cite{morris}
and to map out the phase
diagram as function of temperature and baryon chemical 
potential~\cite{berges1,bj1,braun,pawlow} 
as well as the
calculations of thermodynamic functions~\cite{ipp,bla0} and 
momentum-dependent correlation functions at high 
temperature~\cite{bla1}.

\section{Flow equation and derivative expansion}
\begin{figure}[htb]
\centering
\includegraphics[width=60mm]{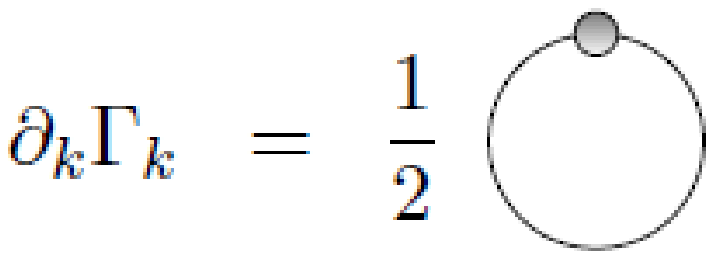}
\caption[Flow]{Diagrammatic representation of the exact flow equation for the
effective action $\Gamma_k[\phi]$. The line denotes the exact field-dependent
propagator and the circle denotes an insertion of the regulator $R_k(q)$.}
\label{glow}
\end{figure}

One implementation of the renormalization group ideas is based
on the effective average action $\Gamma_k[\phi]$
as proposed by Wetterich~\cite{wetterich}.
The effective average action is a functional of a set of fields denoted
by $\phi$ and 
satisfies a flow equation which is an 
integro-differential equation. The subscript $k$ indicates that all
momentum modes $q$ between the UV cutoff of the theory, $\Lambda$, and $k$
have been integrated out. When $k=\Lambda$, no momentum modes have been
integrated out and the effective action is equal to the classical action of
the theory, i.e. $\Gamma_{\Lambda}[\phi]=
S[\phi]$. When $k=0$, all modes have been
integrated out and $\Gamma_0[\phi]$ is equal to the full quantum effective 
action. All quantum and thermal fluctuations from $k=0$
up to $k=\Lambda$ have then been
included.
The exact flow equation for $\Gamma_k[\phi]$ is
\begin{equation}
\partial_k\Gamma_k[\phi]=\frac{1}{2}\mbox{Tr}\Big[\partial_kR_{k}(q)
\big[\Gamma_k^{(2)}+R_k(q)\big]_{q,-q}^{-1}\Big]\;,
\label{flowex}
\end{equation}
where the superscript $n$ on $\Gamma^{(n)}_k[\phi]$ means the $n$'th 
functional derivative of $\Gamma_k[\phi]$
and
the trace is over the spacetime momenta $q$, and indices of 
the inverse propagator matrix. 
The function $R_{k}(q)$ is a regulator and is
introduced in order to implement the renormalization group ideas:
$R_k(q)$
is large for $q<k$ and
small for $q>k$ whenever $0<k<\Lambda$, and $R_{\Lambda}(q)=\infty$.
These properties ensure 
that the modes below $k$ are heavy and decouple, and only
the modes between $k$ and the UV cutoff $\Lambda$ are light and integrated out.
The diagrammatic representation of the flow equation is shown in 
Fig.~\ref{glow}.

The choice of regulator has been discussed extensively in the 
literature~\cite{review5,ball,jensmike,litim,canet}. We employ the regulator 
which is given by
\bqa
R_k(q)&=&(k^2-\mathbf{q}^2)\Theta(k^2-\mathbf{q}^2)\;.
\eqa
This regulator was introduced in Ref.~\cite{litimpaw} 
(see also ~\cite{litim}).
and is particularly convenient in
practical calculations since one can carry out the integral over
three-momenta exactly and this turns
the flow equation into a partial differential equation.

The starting point of our discussion is the classical action $S[\phi]$.
In the present case we consider $N$ complex scalar fields with a quartic
self-interaction and $S[\phi]$ reads
\bqa
S[\phi]&=&\int_{0}^{\beta}d\tau\int_{\bf q}
\left\{
\left(\partial_{\mu}\Phi_i\right)^{\dagger}\left(\partial_{\mu}\Phi_i\right)
+m^2\Phi_i^{\dagger}\Phi_i
-{\sqrt{N\over4}}H\left[\Phi_1^{\dagger}+\Phi_1\right]
+V(\phi)
\right\}\;,
\eqa
where the classical potential is
\bqa
V(\phi)&=&{g^2\over2N}(\Phi_i^{\dagger}\Phi_i)^2\;,
\eqa
and where $i=1,2,3...N$. The complex fields $\Phi_i$ 
are written in terms of real
fields, $\Phi_i=(\phi_{2i-1}+i\phi_{2i})/\sqrt{2}$. We identify $\phi_1$
with $\sigma$, $(\phi_2\pm i\phi_3)/\sqrt{2}$  with $\pi^{\pm}$, and 
$\phi_4,\phi_5,...,\phi_{2N}$ with $2N-3$ neutral $\pi^0$.
The parameter $H$ breaks the $O(2N)$ symmetry explicitly 
to $O(2N-1)$
and 
gives rise to nonzero pion masses. The parameter $m^2$ is negative in the
remainder to ensure spontaneous symmetry breaking in the vacuum
for $H=0$.

A chemical potential $\mu_i$ that corresponds to the conserved charge of the
$U(1)$-symmetry $\Phi_i\rightarrow e^{i\alpha}\Phi_i$ is introduced by
the replacement $\partial_0\Phi_i\rightarrow(\partial_0-\mu_i)\Phi_i$.
The chemical potential acts as the zeroth component of an Abelian
gauge field. The classical action is therefore invariant under the
gauge transformation
\bqa
\label{g1}
\Phi_i&\rightarrow&e^{i\alpha}\Phi_i 
\;,\\ 
\mu_i&\rightarrow&\mu_i+i\partial_0\alpha\;.
\label{g2}
\eqa
The introduction of the isospin chemical potential $\mu_I$ reduces the symmetry
to $O(2)\times O(2N-2)$ if $H=0$, otherwise to $O(2)\times O(2N-3)$.
Chiral symmetry is broken in the vacuum, either spontaneously or explicitly
by a vacuum expectation value $\phi_0$ for $\phi_1$. We also allow for a 
pion condensate by a nonzero value $\rho_0$ 
of $\phi_2$. A pion condensate breaks
the $O(2)$-symmetry. 

After having introduced the isospin chemical potential $\mu_I$,
the classical potential is~\footnote{Note that in the vacuum and
for $H=0$, the potential depends only on combination $\phi_0^2+\rho_0^2$.
One can therefore use the symmetry to rotate away $\rho_0$.
This is in accord with the Vafa-Witten theorem~\cite{waffle} that states that
parity cannot be spontaneously broken in the vacuum.}
\bqa
U_{\Lambda}(\phi)&=&{1\over2}m^2_{\Lambda}(\phi_0^2+\rho_0^2)
-{1\over2}\mu_I^2\rho_0^2+{g_{\Lambda}^2\over8N}
(\phi_0^2+\rho_0^2)^2-\sqrt{N\over2}H\phi_0\;,
\label{boundary}
\eqa
where the subscript $\Lambda$ is 
a reminder that the classical potential equals the
boundary condition for the effective potential $U_k(\phi)$, 
i. e. $U_{\Lambda}(\phi)=V(\phi)$
(see below).

The exact flow equation~(\ref{flowex}) 
cannot not be solved so one must resort to 
approximations. The derivative expansion is a commonly used approximation and
reads
\bqa\nonumber
\Gamma_k[\phi]&=&
\int_0^{\beta}d\tau\int_{\bf q}\left[
Z_k^{1}\left(\nabla\Phi_1\right)^{\dagger}
\cdot\left(\nabla\Phi_1\right)
+Z_k^{2}\left[\left(\partial_0-\mu_I\right)\Phi_1\right]^{\dagger}
\left[\left(\partial_0-\mu_I\right)\Phi_1\right]
\right.\\ &&\left.
+
Z_k^{3}\left(\nabla\Phi_2\right)^{\dagger}\cdot(\nabla\Phi_2)+...
+U_k(\phi)
+...\right]\;,
\eqa
where $Z_k^{i}$ are wavefunction renormalization constants
and $U_k(\phi)$ is the scale-dependent effective potential that depends
on the classical fields that are collectively denoted by $\phi$.
Note that the $Z_k^{i}$'s are in principle different since Lorentz 
invariance
is broken due to both finite temperature and finite density.
The ellipsis indicate higher-order operators including derivatives
such as 
$\Phi^{\dagger}_1\Phi_1\left[(\nabla\Phi_1)^{\dagger}\cdot(\nabla\Phi_1)\right]$. 
In the local-potential
approximation, we set $Z_k^{i}=1$ and omit such higher-order
operators. In this approximation, the flow equation is an integro-differential
equation for 
$U_k(\phi)$~\footnote{Sometimes one further assumes that $U_k(\phi)$ takes a 
polynomial form. The partial differential equation~(\ref{floweq})
is then turned into a
set of coupled differential equations for the couplings.}.

The flow equation for the effective potential $U_k(\phi)$ is derived in the
Appendix and reads
\begin{equation}
\partial_kU_k(\phi)=\frac{4k^dv_{d-1}}{(d-1)}\left\{\bigg[\sum_{i=1}^3
\frac{(1+2n(\sqrt{P_i}))R_i}{2\sqrt{P_i}}\bigg]
+(2N-3)\frac{1+2n(\omega_3)}{2\omega_3}\right\}
\;,
\label{floweq}
\end{equation}
where $d-1$ is the dimension of space, 
$v_d=[2^{d+1}\pi^{d/2}\Gamma(\mbox{$1\over2$})]^{-1}$,
$n(x)=1/(e^{\beta x}-1)$ is the Bose-Einstein distribution function,
$P_i$ and $R_i$ are poles and residues defined in the Appendix.
Moreover, $U_k(\phi)$ is a function of $\rho_1=\frac{1}{2}\phi_0^2$ and 
$\rho_2=\frac{1}{2}\rho_0^2$. We then introduce the shorthand notation
$\partial_1 U_k(\phi)\equiv\partial U_k(\phi)/\partial\rho_1$,
$\partial_2 U_k(\phi)\equiv\partial U_k(\phi)/\partial\rho_2$, and
$\omega_3=\sqrt{k^2+\partial_1U_k}$.


\section{Numerical results and discussion}
In this section, 
we solve the flow equation~(\ref{floweq}) numerically
using a third-order Runge-Kutta method
and discuss the 
resulting
phase diagram.
The boundary potential $U_{\Lambda}(\phi)$ is given by Eq.~(\ref{boundary}),
where we in remainder of this section set $N=2$.
The parameters $m^2_{\Lambda}$ and $g^2_{\Lambda}$ are determined so that 
we reproduce the pion mass $m_{\pi}$, the sigma mass $m_{\sigma}$, and 
the pion decay constant $f_{\pi}$ in the vacuum, i.e. for $T=\mu_I=0$.
In the vacuum we have $\phi_0=f_{\pi}$ and $\rho_0\equiv0$.
The $k$-dependent masses $m_{\pi,k}$ and $m_{\sigma,k}$
can be related
to the effective potential at the $k$-dependent minimum $f_{\pi,k}$
as follows
\bqa
\label{eq:pion-mass-rho}
m_{\pi,k}^2&=&{\partial_1 \tilde{U}_k(\phi)}\bigg|_{\phi_0=f_{\pi,k},\rho_0=0}\\ 
\nonumber
\label{msigmen}
m_{\sigma,k}^2&=&
{\partial_1 \tilde{U}_k(\phi)}\bigg|_{\phi_0=f_{\pi,k},\rho_0=0}
+\phi_0^2{\partial_1^2 \tilde{U}_k(\phi)}\bigg|_{\phi_0=f_{\pi,k},\rho_0=0}\\ 
&=&
m_{\pi,k}^2
+\phi_0^2{\partial_1^2 \tilde{U}_k(\phi)}\bigg|_{\phi_0=f_{\pi,k},\rho_0=0}\;,
\eqa
where $\tilde{U}_k(\phi)=U_k(\phi)+H\phi_0$, i.e. the potential without the 
explicit symmetry breaking term. In particular, the physical pion and sigma 
masses, i.e. the masses for
$k=0$ are given by
\bqa
\label{mpik0}
m_{\pi}^2&=&{\partial_1 \tilde{U}_0(\phi)}\bigg|_{\phi_0=f_{\pi},\rho_0=0}\\
m_{\sigma}^2&=&
m_{\pi}^2
+\phi_0^2{\partial_1^2 \tilde{U}_0(\phi)}\bigg|_{\phi_0=f_{\pi},\rho_0=0}\;.
\eqa
We can now use Eq.~(\ref{msigmen}) together with $\phi_{0,k}=f_{\pi,k}$
to determine the two parameters $m^2_{\Lambda}$ and $g^2_{\Lambda}$.
The fact that $H=f_{\pi}m_{\pi}^2$ implies that this parameter is already known
and the term $-H\phi_0$ can be added to the potential after we have 
determined $m^2_{\Lambda}$ and $g^2_{\Lambda}$.

\begin{figure}[htb]
\centering
\includegraphics[width=110mm]{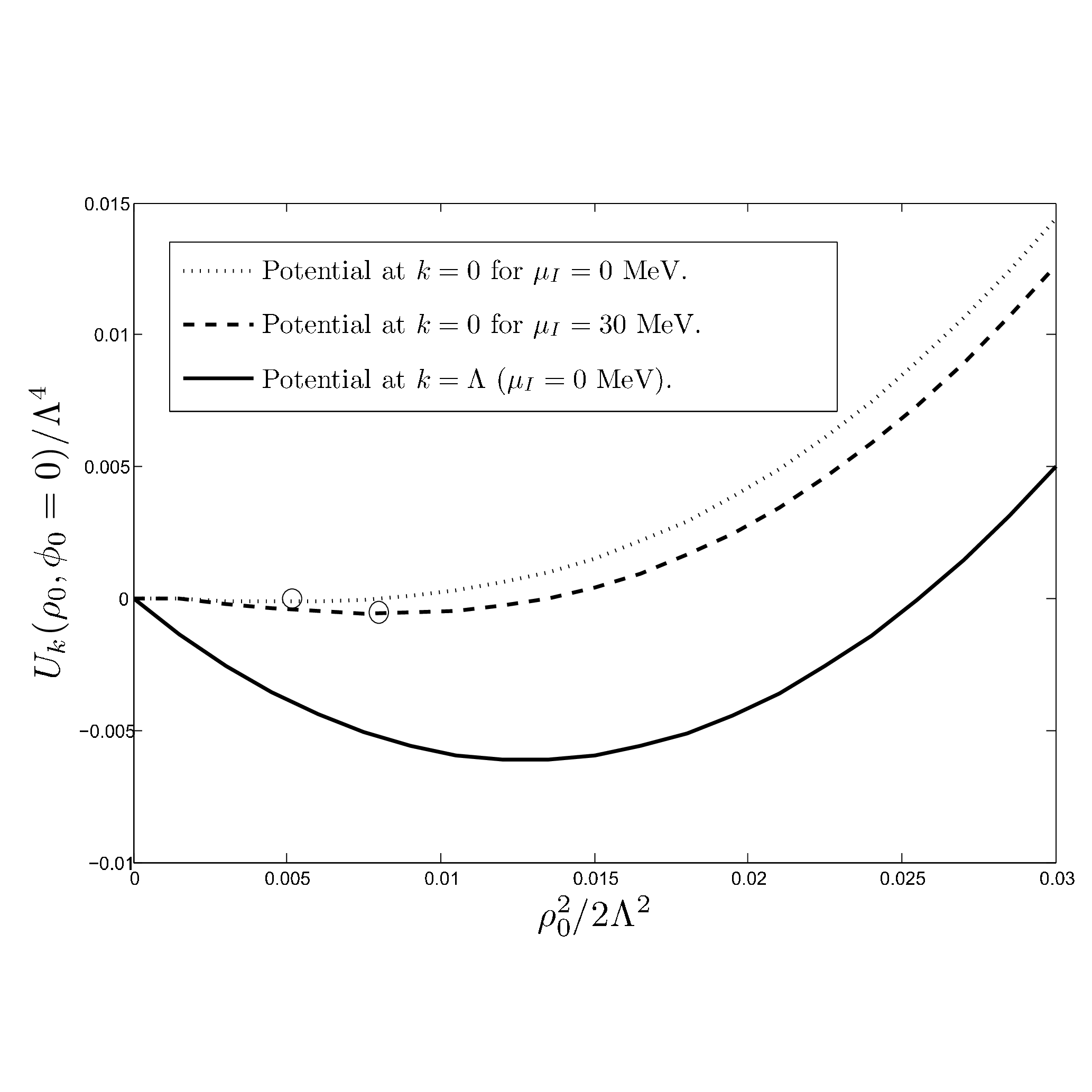}
\caption[Pion-effective potential as a function of $\phi_2$ in the chiral limit]
{The real part of $U_k(\phi_0,\rho_0=0)/\Lambda^4$ as a function of 
$\phi_0^2/2\Lambda^2$ at $T=0$ in the chiral limit. The potential is 
plotted for $k=\Lambda$, and for $k=0$ in the case of $\mu_I=0$  and 
$\mu_I=30$ MeV, where we have neglected the constant added to the potential 
by renormalization. We have used $m_\sigma=400$ MeV and $f_\pi=93$ MeV in order 
to tune the bare parameters of the potential at the cutoff 
$\Lambda=\sqrt{5}m_\sigma$. The minima of the renormalized potentials have 
been marked with open
circles as a guide to the eye.}
\label{fig:plot-pot}
\end{figure}
In Fig.~\ref{fig:plot-pot} we show the real 
part~\footnote{Note that the effective
potential in the broken phase 
has an imaginary part to the left of the minimum.} 
of the 
scaled effective potential as function of $\phi_0^2/2\Lambda^2$ in the 
chiral limit
and for $T=0$ at $\mu_I=0$ and $\mu_I=30$ MeV. 

We
use the experimental value $f_{\pi}=93$ MeV and $m_{\sigma}=400$ MeV to tune the 
bare parameters in this case. We chose the cutoff $\Lambda^2=5m_{\sigma}^2$. This 
gives the values $m^2_{\Lambda}=-0.96\Lambda^2$ and $g_{\Lambda}^2=37.56$. We find 
that the inclusion of a finite chemical potential has the effect of 
increasing the $\rho_0$ value for which $U_k(\phi)$ takes its minimum. We 
also see that including the quantum fluctuations from the bosons
reduces the amount of symmetry breaking. For fermions, the effect is
opposite due a sign change of the vacuum term in the flow equation.
Moreover, if the symmetry
is restored for some values of $T$ and $\mu_I$ at $k=0$, there exists
a scale $k=k_{\rm SSB}$ such $\Gamma[\phi]_k$ still displays 
symmetry breaking when $k>k_{\rm SSB}$.

\begin{figure}[htb]
\centering
\includegraphics[width=110mm]{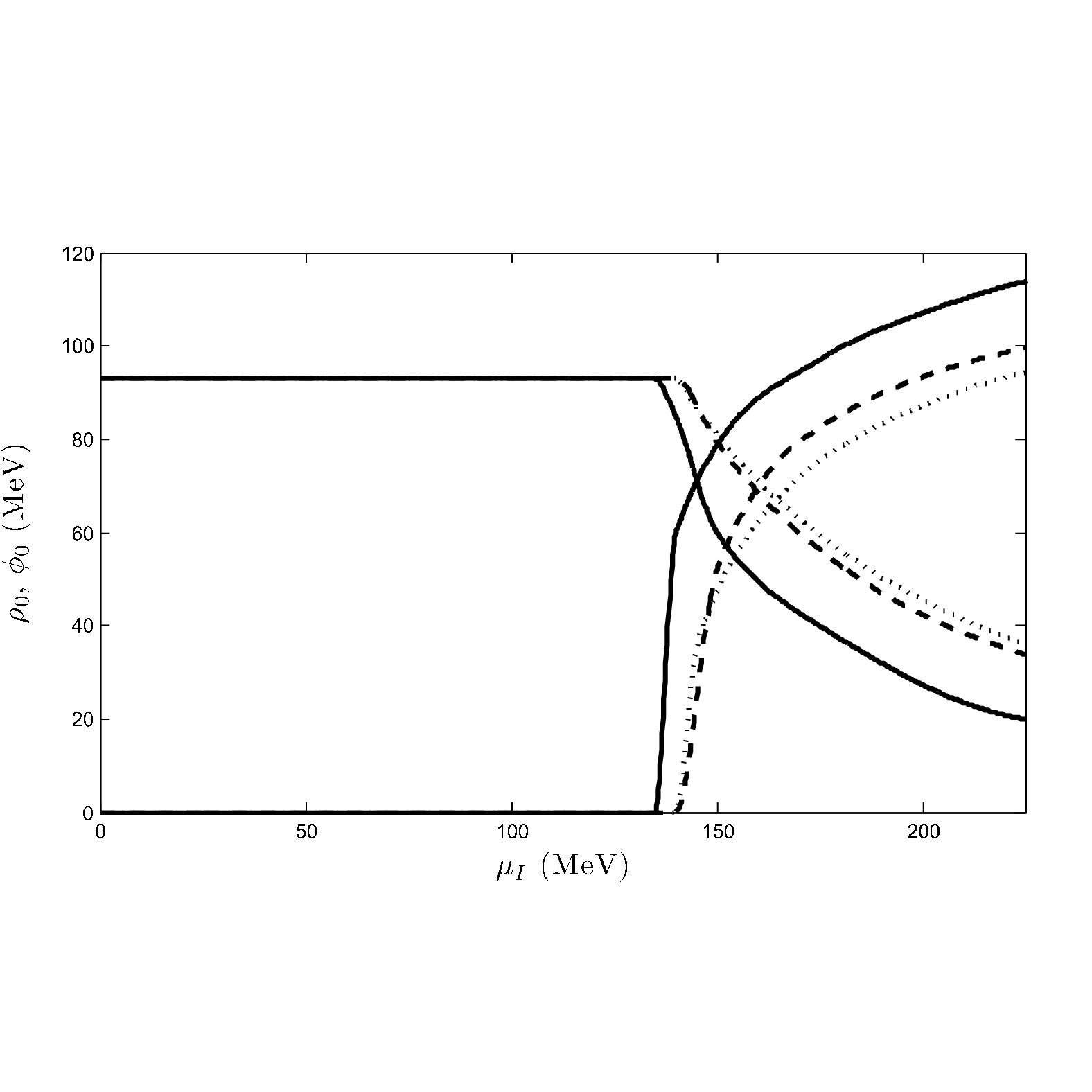}
\caption[Condensate competition]{The chiral condensate and the charged pion 
condensate as a function of isospin chemical potential at the physical point 
at $T=0$. Solid line is the local-potential approximation
with $N=2$, dashed line is the 
large-$N$ limit and the dotted line is the mean-field approximation.
We have used the values
$m_\sigma=600$ MeV, $m_\pi=140$ MeV, and $f_\pi=93$ MeV, and set $\Lambda=600$ 
MeV.}
\label{fig:cond-comp}
\end{figure}

In Fig.~\ref{fig:cond-comp}, we show the chiral condensate 
and pion condensate as function of $\mu_I$ at zero temperature. 
We have used the parameters $f_{\pi}=93$ MeV, $m_{\sigma}=600$ MeV, 
$m_{\pi}=140$ MeV, and $\Lambda=600$ MeV. 
The solid line is the local-potential approximation with $N=2$, the
dashed line is the large-$N$ result, and the dotted line is the mean-field
result. 
In the chiral limit the charged pion condensate is the only condensate present 
at nonzero chemical potential when $T=0$. 
If we go to the physical point however, we see that a chiral condensate 
appears in the direction of $\phi_0$. For small chemical potentials, we find 
that the chiral condensate is the only existing condensate, that is, 
the minimum of $U_k(\phi)$ lies on the $\phi_0$-axis and no charged pion 
condensation takes place. As we increase the chemical potential however, 
the minimum eventually moves away from the $\phi_0$-axis and 
a charged pion condensate is formed, i.e. the chiral condensate is 
(partly) rotated
into a condensate of charged pions. 
At $T=0$ we find that this happens when the chemical potential is about 
the vacuum pion mass. This is in agreement with the results of 
e.g.~\cite{barducci,herpay,heman2,jens1,jenslars1} (but see the discussion
below).

\begin{figure}[htb]
\begin{center}
\includegraphics[width=110mm]{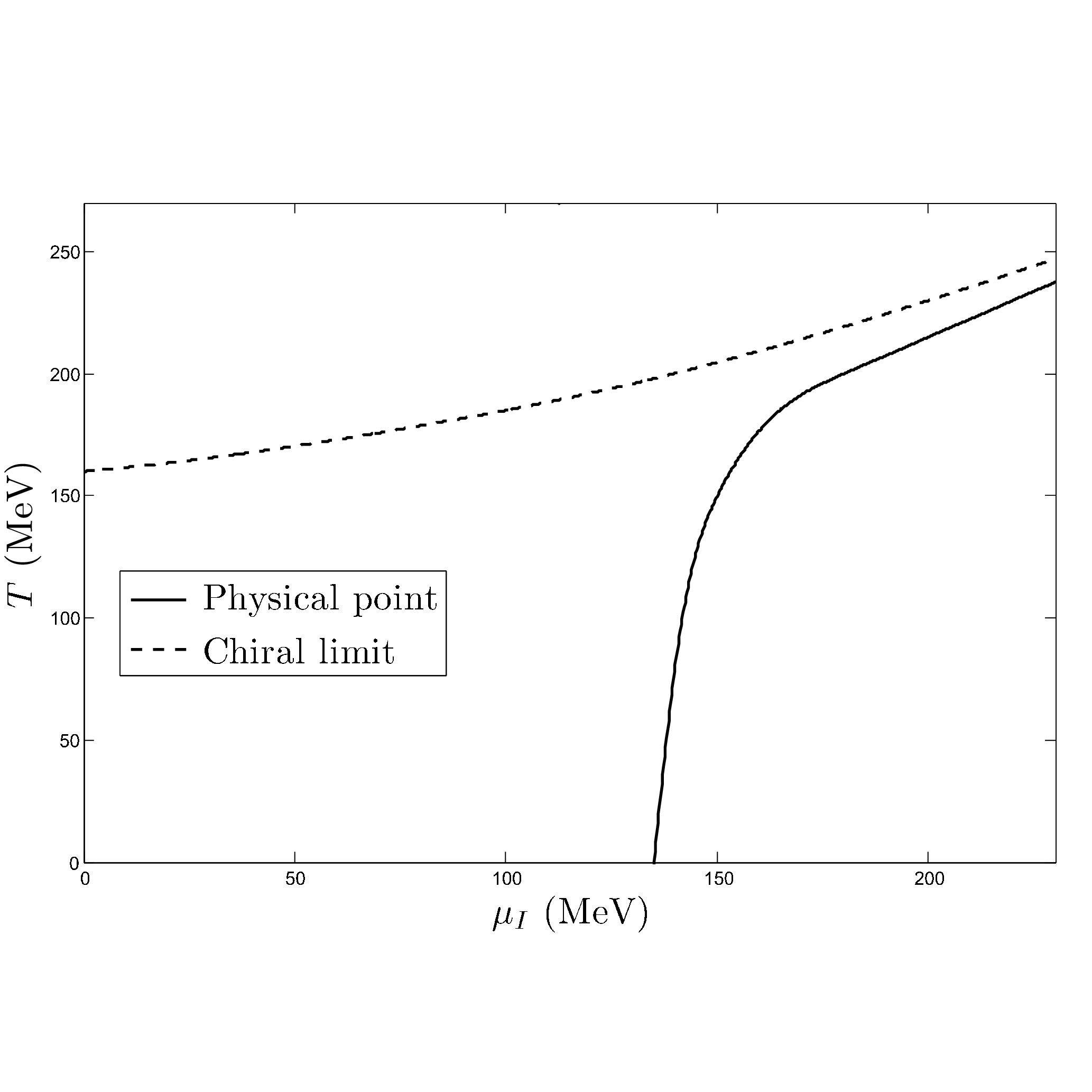}
\caption[Phase diagram; charged pion condensation]{Phase 
diagram for the charged pion condensate in the chiral limit and for the 
physical point. We have chosen physical values $f_\pi=93$ MeV, $m_\sigma=600$ 
MeV and $m_\pi=140$ MeV ($m_\pi=0$ in the chiral limit) in order to tune the 
bare parameters at $T=0$. We have also set $\Lambda=600$ MeV. The symmetric 
phase is above the lines, while the broken phase is below the lines.}
\label{p0}
\end{center}
\end{figure}
In Fig.~\ref{p0}, we plot the phase diagram as function of 
temperature and isospin chemical potential in the chiral limit (dashed line)
and at the physical point (solid line). We use the same parameters as 
before.
The phase diagram is in qualitative agreement
with previous results~\cite{herpay,heman2,jens1,jenslars1}. 
In particular, the phase transition to a condensate of pions is
of second order and this is in accord with universality-class arguments.
Note, however, that the critical chemical potential for $T=0$ 
and at the physical point is different from the pion mass $m_{\pi}$ and this 
is in 
disagreement with exact results.
In fact, it can be shown that the critical chemical potential 
for Bose condensation at $T=0$ is equal to the mass of the condensing mode.
Using the bare 
parameters we found at $T=\mu_I=0$, we see from 
Fig.~\ref{p0} that we obtain a slightly different 
value for the critical chemical potential, namely $\mu_I^c=135$ MeV. The 
reason we do not get the pion mass is that we are working in the 
local-potential approximation, 
where wave-function renormalization is neglected. Recall that to include an 
isospin chemical potential we should let 
$\partial_0\Phi\rightarrow(\partial_0-\mu_I)\Phi$, where we view the 
Lagrangian as a function of complex field variables. This term remains 
unrenormalized in the local-potential approximation
 and so the effective action is not gauge invariant
under the transformations~(\ref{g1}) -~(\ref{g2}).
This is in contrast to the full quantum effective action
at $T=0$~\cite{son}.

A simple solution to this problem is to first solve the flow equation for 
the effective potential $U_k(\phi)$ in the absence of the isospin chemical
potential and then gauge the effective action in the local-potential
approximation
by the substitution $\partial_0\Phi\rightarrow(\partial_0-\mu_I)\Phi$.
The effective potential at scale $k$ may then be written as
\begin{equation}
U_{k}(\phi)=\tilde{U}_k(\phi)-{1\over2}\mu_I^2\rho_0^2-H\phi_0,
\label{o4def}
\end{equation}
where $\tilde{U}_k(\phi)$ denotes the 
$O(4)$-symmetric potential, i.e. the potential satisfying Eq.~(\ref{floweq})
with $\mu_I=0$, that is Eq. (\ref{eq:final-RGeq-nocp}) in the Appendix. We call
this the $O(4)$-approximation.
In order to get a charged pion 
condensate at 
$k=0$, one needs a negative partial
derivative in the $\rho_0$ direction at the minimum
of $U_k(\phi)$. When $\mu_I$ is 
just large enough for this to happen, we obtain
\bqa\nonumber
0&=&\partial_2U_{k=0}(\phi)\\
&=&\partial_2\tilde{U}_k(\phi)-\mu_I^2.
\label{o4p}
\eqa
Using the $O(4)$-invariance of $\tilde{U}_k(\phi)$ and 
combining Eq.~(\ref{o4p}) with Eq.~(\ref{mpik0}), this yields a critical 
chemical potential 
\begin{equation}
\mu_I^c=m_{\pi}\;.
\label{eq:mu=pi}
\end{equation}
While the effective potential $U_k(\phi)$ defined by 
Eq.~(\ref{o4def}) is in agreement with exact results, its dependence on the
isospin chemical potential is of course trivial.

The large-$N$ limit provides us with another 
flow equation for $U_k(\phi)$ which is 
independent of the chemical potential. It is given by Eq.~(\ref{flowln})
in the Appendix:
\bqa
\partial_kU_k=
\frac{4k^dv_{d-1}}{(d-1)\omega_3}\left[1+2n(\omega_3)\right]\;.
\label{flowl0}
\eqa
Thus using the same arguments as above, 
we obtain the critical chemical potential for pion condensation
which is equal to the pion mass.

\begin{figure}[htb]
\centering
\includegraphics[width=110mm]{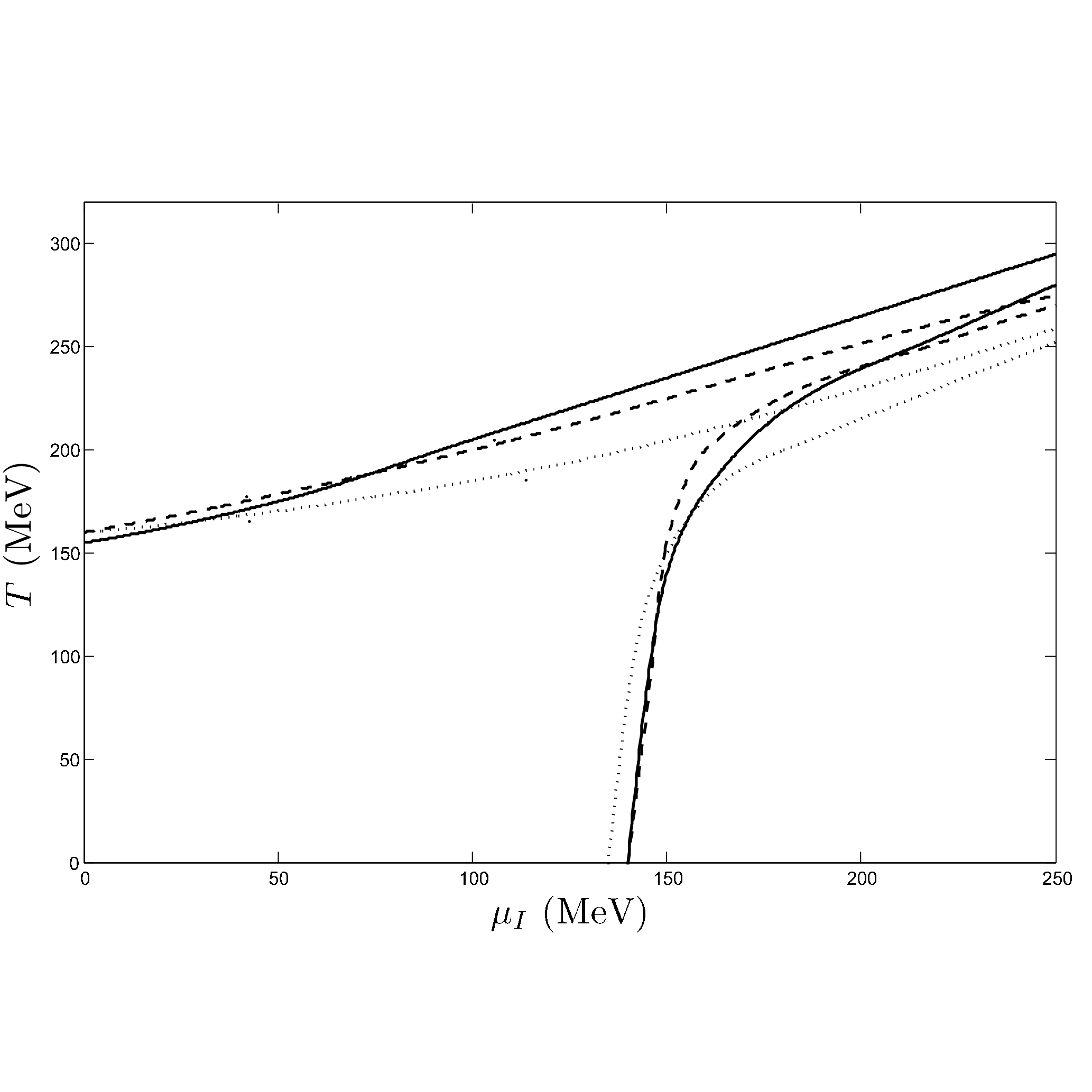}
\caption[Phase diagram; Charged pion condensate, $O(4)$-approximation and 
the large-$N$ limit]{The phase diagram for the $O(4)$-approximation 
(solid line) and the large-$N$ limit (dashed line), where we have set 
$m_\pi=140$ MeV, $f_\pi=93$ MeV, $m_\sigma=600$ MeV, and the 
cutoff $\Lambda=1500$ MeV to determine the parameters 
$m_{\Lambda}^2$ and $g_{\Lambda}$. We have also 
plotted the previous obtained phase diagram for comparison (dotted line).}
\label{fig:Large-N-limit}
\end{figure}

In Fig.~\ref{fig:Large-N-limit}, we show the phase diagram obtained by 
solving the flow equation for $U_k(\phi)$ in the $O(4)$-approximation and in the
large-$N$ limit. For comparison, we also show the phase diagram obtained
by solving Eq.~(\ref{floweq}) and was shown in Fig.~\ref{p0} albeit with
a different
cutoff $\Lambda$. In this plot, we used $\Lambda=1500$ MeV
(Note that the dotted lines are from Fig.~\ref{p0} with $\Lambda=600$ MeV).
We would like to point out that both the $O(4)$ approximation
and the large-$N$ approximation yield a second-order transition  and adding
the term $-\mbox{$1\over2$}\mu_I\rho_0^2$ does not change this.
Hence all three approximations predict the correct order of the transition.

In closing, we remark that it would be interesting to include the effects
of wavefunction renormalization. This would lead to a set of coupled equations
for $Z_k^{i}$ and $U_k(\phi)$. Hopefully it would close the gap between
the value $\mu_I^c=135$ MeV obtained in this work and the exact result
$\mu_I^c=m_{\pi}=140$ MeV and so perhaps provide a practical solution to the 
problems
of the local-potential approximation.
Similarly, it would be interesting to apply the method to the chiral quark
model. Since this model has quark degrees of freedom, 
one can simultaneously study the effects of a baryon and an isospin 
chemical potential. In particular, one can investigate the BEC-BCS
transition as one increases $\mu_I$. For small values of $\mu_I$, but
larger than $m_{\pi}$, there is a condensate of pions.
For asymptotically large values of $\mu_I$, and therefore large densities,
the system is weakly coupled due to asymptotic freedom. 
In that case, the weak attraction gives rise to Cooper pairs and a BCS
state. The order parameter has the same quantum numbers as the BEC state
and so the transition is a crossover~\cite{iso}.

The BEC-BCS transition has been discussed at length in the functional
renormalization group framework for ultracold atomic 
gases~\cite{func0,func1,func2}. The crossover in 
these nonrelativistic systems is realized by changing the $s$-wave scattering
length. This is done by varying an external magnetic field $B$
and exploiting the existence of a Feshbach resonance at $B_0=834$ Gauss.
In this way one can basically change the coupling constant from minus infinity
to plus infinity. A BEC state is obtained for positive scattering lengths and
a BCS state for negative scattering lengths, i. e. on each side of the
resonance.


\section*{Acknowledgments}
The authors would like to thank Tomas Brauner,
Holger Gies, and Michael Strickland for useful discussions.


\section{Appendix}
In this Appendix, we derive the flow equation~(\ref{floweq}). 
The starting point is the exact flow equation~(\ref{flowex})~\cite{wetterich}
\bqa
\partial_k\Gamma_k[\phi]=\frac{1}{2}\mbox{Tr}
\Big[\partial_kR_{k,q}\big[\Gamma_k^{(2)}+R_k(q)\big]_{q,-q}^{-1}\Big]\;.
\label{exact2}
\eqa
The effective potential $U_k(\phi)$ is defined by evaluating the
effective action $\Gamma_k[\phi]$ for space-time independent values of the
fields, i. e.
\bqa
U_k(\phi_{\rm uni})&=&{1\over VT}\Gamma_k[\phi_{\rm uni}]\;,
\label{udef}
\eqa
where $VT$ is the spacetime volume of the system and $\phi_{\rm uni}$
is constant.
Taking the derivative of Eq.~(\ref{udef}) with respect to $k$, we obtain
the RG-equation for the effective potential
\begin{equation}
\partial_kU_k(\phi_{\rm uni})=\frac{1}{2VT}\mbox{Tr}\bigg[\partial_kR_{k,q}
\big[\Gamma^{(2)}_k[\phi_{\rm uni}]+R_k(q)\big]^{-1}_{q,-q}\bigg].
\label{eq:RG-pot}
\end{equation}
In the remainder we drop the subscript ${\rm uni}$.
With $\Gamma_k[\phi]$ given as in Eq.~(\ref{exact2}), 
we find the Fourier transform of $\Gamma^{(2)}_k[\phi]$ 
to be
\begin{equation}
\Gamma^{(2)}_{k,i,j,q,q'}=(2\pi)^{d}\delta(q+q')
\Bigg[\frac{\partial^2U_k}{\partial\phi_i\partial \phi_j}
+\delta_{ij}q^2-2\mu_Iq_0(\delta_{j2}\delta_{i3}-\delta_{i2}\delta_{j3})\Bigg]\;,
\label{eq:GFE-2}
\end{equation}
where $d$ denotes the spacetime dimension and $i,j$ are the indices of the
propagator.
Note that by the presence of the chemical potential, 
$\Gamma^{(2)}_{k,i,j,q,q'}$
is not a 
diagonal matrix. Also note that we cannot rotate the field vector 
$\phi=(\phi_1,..,\phi_{2N})$ to point in one specific direction as we do not 
have $O(2N)$ symmetry. This has been explicitly broken to a 
$O(2)\times O(2N-2)$ by the isospin chemical potential $\mu_I$. The best we 
can do is to write $\phi=(\phi_0,0,\rho_0,,0,..,0)$ to incorporate both 
a chiral condensate and a charged pion condensate.
This will also give rise to off-diagonal 
entries in the matrix. We now set $N=2$ in order to simplify the derivation. 
Returning to a general $N$ is straightforward. 

We may assume, 
because of the $O(2)\times O(2)$ symmetry, 
that $U_k(\phi)=U_k(\rho_1,\rho_2)$ where 
\bqa\nonumber
\rho_1&=&\frac{1}{2}\langle\phi_1^2+\phi_4^2\rangle \\
&=&\frac{1}{2}\phi_0^2\;,
\\ \nonumber
\rho_2&=&\frac{1}{2}\langle\phi_2^2+\phi_3^2\rangle
\\
&=&\frac{1}{2}\rho_0^2\;.
\eqa
If we denote the fields by double 
indices $\phi=(\phi_{11},\phi_{12},\phi_{21},\phi_{22})$, we can derive the 
identity
\begin{equation}
\frac{\partial^2U_k}{\partial\phi_{ai}\partial\phi_{bj}}
=\frac{\partial U_k}{\partial\rho_b}\delta_{ab}\delta_{ij}+\phi_{ai}\phi_{bj}
\frac{\partial^2U_k}{\partial\rho_a\partial\rho_b}.
\end{equation}
The Fourier transform of 
$R_k(x-y)$ is $(2\pi)^d\delta(q+q')R_k(q)=R_{k,q,q'}$. If we add 
$\delta_{ij}R_{k,q,q'}$ to $\Gamma^{(2)}_{k,i,j,q,q'}$, we obtain the matrix
\bqa\nonumber
&\big[\Gamma^{(2)}_k+R_k\big]_{q,q'}=(2\pi)^d\delta(q+q')\times \\
&\left(\begin{array}{cccc}\partial_1U_k+2\rho_1\partial_1^2U_k+F_k(q)&0&2\sqrt{\rho_1\rho_2}\partial_1\partial_2U_k&0\\
0&\partial_2U_k+F_k(q)&-2\mu_Iq_0&0\\
2\sqrt{\rho_1\rho_2}\partial_1\partial_2U_k&2\mu_Iq_0&\partial_2U_k+2\rho_2\partial_2^2U_k+F_k(q)&0\\
0&0&0&\partial_1U_k+F_k(q)\end{array}\right),
\label{eq:Gamma2-matrix}
\eqa
where $F_k(q)=q^2+R_k(q)$. 
Using the fact that the inverse in Fourier space satisfies
\begin{equation}
\int_{q'}F(q_1,q')_{ik}F^{-1}(q_2,q')_{kj}=(2\pi)^d\delta(q_1-q_2)\delta_{ij}\;,
\label{eq:fourier-inverse}
\end{equation}
we may invert $\big[\Gamma^{(2)}_k+R_k\big]_{q,q'}$ to obtain the full 
$k$-dependent propagator. This may then be inserted into Eq. (\ref{eq:RG-pot}) 
in order to obtain the RG equation for the effective potential. 
Going to imaginary time we replace the integral over $q_0$ by a Matsubara sum 
$T\sum_n$. This yields
\bqa\nonumber
\partial_kU_k&=&\frac{T}{2}\sum_n\int_\mathbf{q}\partial_kR_k
\Bigg[\bigg(2(q^2+R_k+\partial_1U_k+2\rho_1\partial_1^2U_k)
(q^2+R_k+\partial_2U_k+\rho_2\partial_2^2U_k)\\ \nonumber
&&-4\rho_1\rho_2(\partial_1\partial_2U_k)^2+(q^2+R_k+\partial_2U_k)
(q^2+R_k+2\rho_2\partial_2^2U_k)+4\mu_I^2\omega_n^2\bigg)\bigg/\\ \nonumber
& &\bigg((q^2+R_k+\partial_1U_k+2\rho_1\partial_1^2U_k)
\Big[(q^2+R_k+\partial_2U_k+2\rho_2\partial_2^2U_k)(q^2+R_k+\partial_2U_k)
+4\mu_I^2\omega_n^2\Big]\\
&&-4\rho_1\rho_2(\partial_1\partial_2U_k)^2(q^2+R_k+\partial_2U_k)\bigg)+
\frac{1}{q^2+R_k+\partial_1U_k}\Bigg],
\label{eq:full-eq-cp}
\eqa
where now $q^2=\omega_n^2+\mathbf{q}^2$ and $\omega_n=2\pi nT$ denotes
the Matsubara frequencies. The difference between this and arbitrary $N$ is 
that there would be $2N-3$ propagators of the form of the last term in the 
above expression.

Performing the integral over $\mathbf{q}$ and returning to arbitrary $N$, we 
may simplify this by writing
\begin{equation}
\partial_kU_k=\frac{4Tk^dv_{d-1}}{(d-1)}\sum_n\bigg[\frac{N(\omega_n^2)}
{D(\omega_n^2)}+\frac{2N-3}{\omega_n^2+k^2+\partial_1U_k}\bigg]\;,
\label{eq:RGeq-contour-form}
\end{equation}
where we have defined 
\bqa\nonumber
N(\omega_n^2)&=2(k^2+\omega_n^2+\partial_1U_k+2\rho_1\partial_1^2U_k)
(k^2+\omega_n^2+\partial_2U_k+\rho_2\partial_2^2U_k)-4\rho_1\rho_2
(\partial_1\partial_2U_k)^2\\
&+(k^2+\omega_n^2+\partial_2U_k)
(k^2+\omega_n^2+\partial_2U_k+2\rho_2\partial_2^2U_k)+4\mu_I^2\omega_n^2\;,
\\ \nonumber
D(\omega_n^2)&=(\omega_n^2+k^2+\partial_1U_k+2\rho_1\partial_1^2U_k)
\Big[(k^2+\omega_n^2+\partial_2U_k+2\rho_2\partial_2^2U_k)
(k^2+\omega_n^2+\partial_2U_k)\\
&+4\mu_I^2\omega_n^2\Big]+4\rho_1\rho_2(\partial_1\partial_2U_k)^2
(k^2+\omega_n^2+\partial_2U_k)\;.
\eqa
The denominator $D(\omega_n^2)$ is a third-degree polynomial in $\omega_n^2$.
Writing out its roots $P_i$ 
would take several pages. However, they can in principle 
be found, so we may 
for the sake of argument assume that we have found them. 
If we denote by $R_i$ the residue of the three poles $P_i$ of $1/D(\omega_n^2)$,
we may write
\begin{equation}
\frac{N(\omega_n^2)}{D(\omega_n^2)}=\sum_{i=1}^3\frac{R_i}{\omega_n^2+P_i}.
\label{eq:fraction}
\end{equation}
Summing over the Matsubara frequencies yields
\begin{equation}
\partial_kU_k=\frac{4k^dv_{d-1}}{(d-1)}
\left\{\sum_{i=1}^3\bigg[
\frac{(1+2n(\sqrt{P_i}))R_i}{2\sqrt{P_i}}\bigg]
+(2N-3)\frac{1+2n(\omega_3)}{2\omega_3}\right\}\;.
\label{eq:final-RGeq-cp}
\end{equation}
Finding the poles $P_i$ and the residues $R_i$ is done numerically, by use of 
the residue function in Matlab.

In the case of zero chemical potential there is only the possibility of a 
chiral condensate. The situation is again $O(2N)$ symmetric.
Eq.~(\ref{eq:fraction}) then reduces to
\begin{equation}
\frac{N(\omega_n^2)}{D(\omega_n^2)}=
\frac{1}{\omega_n^2+k^2+\partial_1U_k(\rho_1)+2\rho_1\partial_1^2U_k(\rho_1)}
+\frac{1}{\omega_n^2+k^2+\partial_1U_k(\rho_1)},
\end{equation}
The equation thus obtained is
\begin{equation}
\partial_kU_k=\frac{4k^dv_{d-1}}{(d-1)}
\left\{\frac{1+2n(\omega_1)}{2\omega_1}
+(2N-1)\frac{1+2n(\omega_2)}{2\omega_2}\right\}\;,
\label{eq:final-RGeq-nocp}
\end{equation}
where $\omega_1=\sqrt{k^2+\partial_1U_k(\rho_1)+2\rho_1\partial^2U_k(\rho_1)}$ 
and $\omega_2=\sqrt{k^2+\partial_1U_k(\rho_1)}$.

The large-$N$ limit of Eq.~(\ref{eq:final-RGeq-cp}) 
is obtained by
scaling the fields $\phi_i\rightarrow\sqrt{N}\phi_i$. This implies that
$U_k\rightarrow NU_k$ on the left-hand side of Eq.~(\ref{eq:final-RGeq-cp}).
Letting $N\rightarrow\infty$, we obtain
\bqa
\partial_kU_k=
\frac{4k^dv_{d-1}}{(d-1)\omega_3}
\left[1+2n(\omega_3)\right]\;.
\label{flowln}
\eqa
Note that the flow equation~(\ref{flowln}) is independent of the 
isospin chemical potential.

\end{document}